\documentclass{aa}  
\usepackage{xcolor}

\usepackage{graphicx}
\usepackage{txfonts}
\usepackage{lipsum}
\usepackage{subcaption}         
\usepackage{lscape}             
\usepackage{placeins}           
                    
\usepackage{float}
\usepackage[colorlinks = true,
            linkcolor = blue,
            urlcolor  = blue,
            citecolor = blue,
            anchorcolor = blue]{hyperref}

\newcommand{\poet}{PoET}
\newcommand{\espresso}{ESPRESSO}
\newcommand{\shabar}{SHABAR}

\usepackage{orcidlink}

\begin{document}

   \title{Daytime seeing variations in Paranal through a SHABAR system}

   \subtitle{Description of the method and first measurements from the PoET solar telescope}

%
%
%

\author{A. M. Silva\inst{\ref{inst1}, \ref{inst2}} \orcidlink{0000-0003-4920-738X}
   \and B. Wehbe \inst{\ref{inst3}, \ref{inst15}}  \orcidlink{0000-0002-9433-871X}
   \and A. Oliveira  \inst{\ref{inst3}, \ref{inst15}}  
   \and M. Abreu \inst{\ref{inst3}, \ref{inst15}}  
   \and C. Cardenas-Conde \inst{\ref{inst1}, \ref{inst16}}
   \and A. Cabral  \inst{\ref{inst3}, \ref{inst15}}  
   \and P. Moreno \inst{\ref{inst1}}
   \and J. H. C. Martins \inst{\ref{inst1}}
   \and M. A. Monteiro \inst{\ref{inst1}}
   \and N. C. Santos \inst{\ref{inst1}, \ref{inst2}}
   \and R. Gafeira \inst{\ref{inst17}, \ref{inst18}}
   }

\institute{
   Instituto de Astrof\'{\i}sica e Ci\^encias do Espa\c{c}o, CAUP, Universidade do Porto, Rua das Estrelas, 4150-762 Porto, Portugal \label{inst1}
   \and Departamento de F\'{\i}sica e Astronomia, Faculdade de Ci\^encias, Universidade do Porto, Rua do Campo Alegre, 4169-007 Porto, Portugal \label{inst2} 
   \and Instituto de Astrof\'{\i}sica e Ci\^encias do Espa\c{c}o, Universidade de Lisboa, Campo Grande, 1749-016 Lisboa, Portugal\label{inst3}
   \and Departamento de F\'{\i}sica, Faculdade de Ciências, Universidade de Lisboa, Campo Grande, P-1749-016 Lisboa, Portugal \label{inst15}
   \and  University of North Carolina, Chapel Hill, United States \label{inst16}
   \and Instituto de Astrof\'{i}sica e Ci\^{e}ncias do Espa\c{c}o, Department of Physics, University of Coimbra, Rua Larga, 3004-516 Coimbra, Portugal \label{inst17}
   \and Geophysical and Astronomical Observatory, Faculty of Science and Technology, University of Coimbra, Rua do Observatório s/n, 3040-004 Coimbra, Portugal \label{inst18}
   }

   \date{Received September 30, 20XX}


   \abstract
   {}
   {In this paper we present the software and first seeing measurements from a  SHAdow Band Ranger (\shabar) system that is coupled with the recently comissioned \poet{} solar telescope, installed at ESO's VLT platform in the Paranal Observatory. The seeing measurements that are collected by this system play a pivotal role in the observational strategy of \poet, as it employs apertures ranging between 1 and 55 arcseconds to observe resolved solar regions. As such, the usage of the smaller apertures hinges upon good observational conditions, which are assessed through real-time measurements from the instrument.}
   {We employ a \shabar{} system, consisting of a series of scintillometers that are non-uniformly spaced in a bar, measuring the scintillation of the incoming light. The scintillation is measured in individual detectors, which are then used to characterize the atmospheric turbulence at a given range of heights. We use previously derived formulations, following Kolmogorov turbulence models, to model the collected data and transform it into a seeing measurement.}
   {We find that the Paranal daytime seeing can reach values close to 1 arcsecond in the early morning, before the ground temperature increases and creates turbulence in the lower layers of the atmosphere. Over the day, the seeing gradually worsens, with the median seeing reaching values larger than 4 arcseconds in the afternoon. We also find that these variations appears to be stable over the first weeks of observations, yielding compatible results when comparing 37 independent days of operations, with a median seeing of 2.37 arcseconds.}
   {}


   \maketitle
   \nolinenumbers

\section{Introduction}

One of the largest goals of current day astrophysics is the detection and characterization of Earth-2.0, a rocky planet with the physical conditions to hold liquid water in its surface, orbiting a Sun-like star. For such purpose, one of the  major workhorses lies in the Radial Velocity (RV) technique, based on measuring variations in a star's position along the line of sight. This is typically done through measuring the displacement of spectral lines, due to the Doppler effect, with ground-based spectrographs. Advances in instrumentation and data analysis have pushed the noise floor of current instruments \citep{jurgensonEXPRESNextGeneration2016,pepeESPRESSOVLTOnsky2021,seifahrtMAROONXFirstTwo2022} to values close to the expected RV signature of an Earth-twin ($\sim$ 9 $m\ s^{-1}$). However, we are yet to detect such planets, mainly due to the spurious signals that are introduced from the host stars, introducing time- and wavelength-dependent contamination \citep[e.g.,][]{costesLongtermStellarActivity2021,dumusquePlanetaryDetectionLimits2011,dumusqueMeasuringPreciseRadial2018,almoullaMeasuringPreciseRadial2022,fariaCandidateShortperiodSubEarth2022}. This has motivated the construction of multiple solar telescopes \citep{dumusqueThreeYearsHARPSN2021,farretjentinkABORASPolarimetric10cm2022,linObservingSunStar2022,rubenzahlStaringSunKeck2023,santosPoETParanalSolar2025}, with the end goal of understanding the Sun to then transpose this knowledge to better interpret the variability present in the data of other stars. However, until recently, no existing instrument allows the study of resolved regions of the solar surface across large wavelength intervals.

With this in mind, the recently comissioned Paranal solar Espresso Telescope \citep[\poet][]{leitePoETParanalSolar2024, santosPoETParanalSolar2025} had its first light in April 2026. It is connected to the \espresso{} \citep{pepeESPRESSOVLTOnsky2021} spectrograph, installed at ESO's Very Large Telescope (VLT). It is designed to both observe in a Sun-as-a-Star mode (integrated solar disk, with simultaneous wavelength calibration) or to inject light from resolved regions in the solar disk, with on-sky apertures ranging from 1 to 55 arcseconds. We refer to \cite{santosPoETParanalSolar2025} for a detailed description of the system. The range of apertures, coupled with the ultra-high resolution of \espresso{} (R>200000) and the ability to simultaneously observe the full optical spectrum (380-780 nm) makes \poet{} a unique instrument to study the Sun and the different sources of stellar "noise" in exoplanet research. However, its usage is not without its challenges, particularly when observing in the resolved mode, where atmospheric seeing can degrade image quality. 

Recently, \cite{griffithsComparisonNextgenerationTurbulence2024} employed a daytime seeing SHIMM \citep{griffithsDemonstrating24hourContinuous2023} device to measure seeing in the Paranal platform, reporting a 2.65 median arcsecond seeing over a period of a few days, larger than \poet's smallest on-sky aperture. As such, it is important to have a real-time assessment of the seeing conditions, to allow the observers to take informed decisions on which aperture can be used at any given time. This motivated the construction of a SHAdow Band Ranger \citep[\shabar][]{beckersSeeingMonitorSolar2001, sliepenSeeingMeasurementsAutonomous2010, wehbeImplementationSeeingMeasurement2024,wehbeSeeingMeasurementDevice2026} system, attached to the telescope and providing real time seeing measurements.

In this manuscript we present the methodology that is used to analyze the data from the \shabar{} system, and daytime seeing measurements collected during the first month of operations of \poet. In Section \ref{sect:shabar_hardware_description} we describe the hardware  of the \shabar{} system, with Section \ref{sect:shabar_data_analysis} describing the data analysis and processing workflow. In Section \ref{Sec:paranal_daytime_seeing} we present the daytime seeing measurements collected during the first month of operations of \poet. Finally, in Section \ref{Sec:conclusions} we present our conclusions and future work.

\section{The SHABAR system} \label{sect:shabar_hardware_description}

   The \shabar{} system consists in a series of scintillometers that are non-uniformly spaced in a bar, as per Table \ref{Tab:detector_separation}, measuring the scintillation of the incoming light and providing non-redundant baselines for height characterization. We refer to \citet{wehbeSeeingMeasurementDevice2026} for a detailed description of the hardware, alongside a comparison with another \shabar{} system that is installed in the Swedish Solar Telescope (SST) at the Roque de los Muchachos Observatory, La Palma in the Canary Islands \citep{scharmer1meterSwedishSolar2003}. It is also important to note that the previous results from \citep{wehbeSeeingMeasurementDevice2026} revealed a good agreement between the two instruments, validating our processing and analysis workflow that will be described later in this manuscript.

   \begin{table}
      \caption{Physical position of the detectors in the \shabar{} system, in millimeters.}             
      \label{Tab:detector_separation}
      \centering                     
      \begin{tabular}{c|c}          
      \hline\hline                   
      Detector & Position in bar [mm]\\    
      \hline                               
         0 & 0 \\  
         1 & 20 \\
         2 & 55.2 \\
         3 & 117.7 \\
         4 & 228.1 \\
         5 & 432.4 \\               
      \end{tabular}
   \end{table}

\begin{figure}[h]
      \centering
      \resizebox{\hsize}{!}{\includegraphics{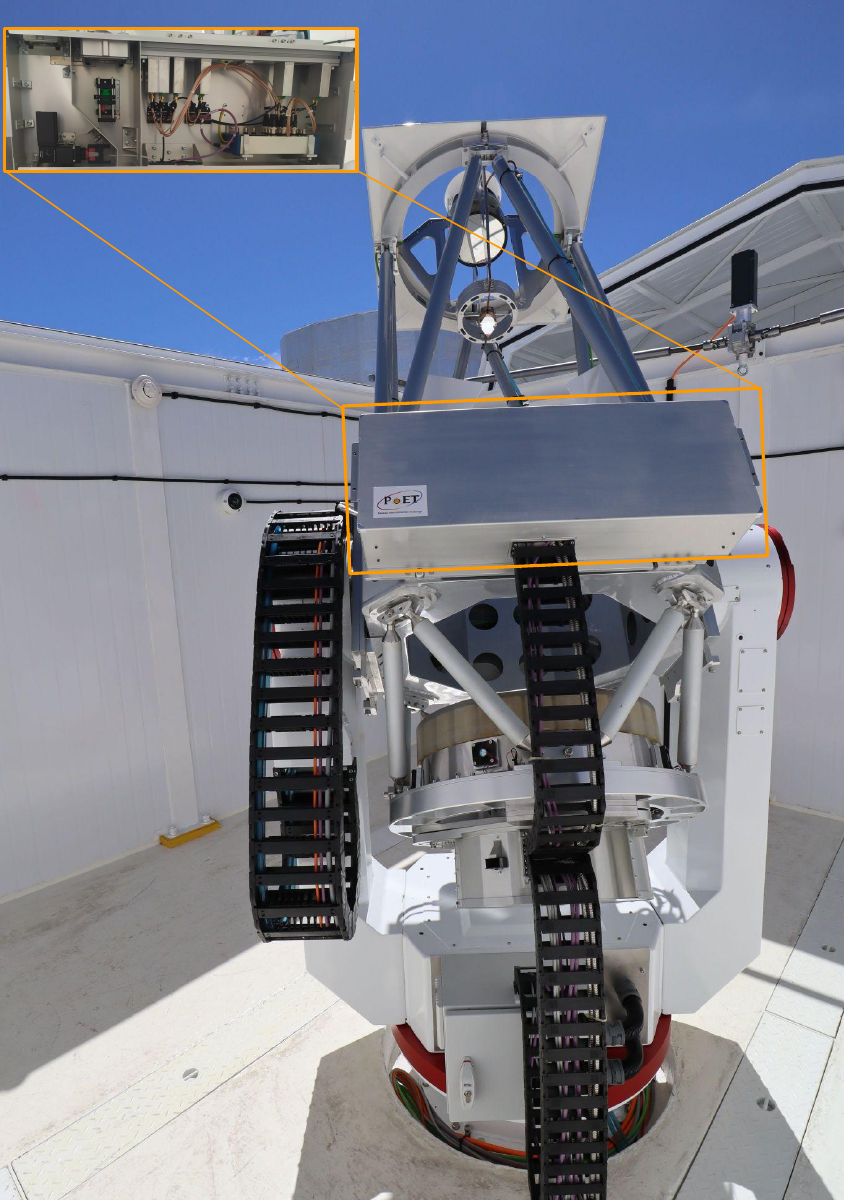}}
      \caption{Photo of the \poet{} solar telescope, with a piggyback pointing telescope, highlighted in orange. A picture of the internal components of the pointing telescope is also shown, with the \shabar{} system (scintillometers and DAQ) on the right side of the box. The hardware is fully described within \citet{wehbeSeeingMeasurementDevice2026}, to which we refer for further details.}
      \label{Fig:poet_picture}
   \end{figure}

   The scintillation, measured in the individual detectors is caused by optical propagation of distorted wavefronts, and can be used to infer the seeing at different altitudes. The system is attached to the telescope and aligned with it, leading to no seeing measurements during downtime of the telescope or times at which it is not pointing at the Sun. Figure \ref{Fig:poet_picture} shows a picture of the \poet{} telescope, highlighting the encasing of the \shabar{} system.

   The individual scintillometers collect the AC and DC signals at a rate of 1000 Hz, in intervals of 2 seconds, and with a voltage range of $\pm$ 9 V. The interval was selected to be similar to the \citet{sliepenSeeingMeasurementsAutonomous2010}, ensuring that we are sensitive to seeing variations in short timescales. These measurements will then be translated into seeing (as per the formulations of Section \ref{sect:shabar_data_analysis}). Since the Sun is only visible at zenithal angles smaller than 60 degrees, the \shabar{} system will collect daily data for up to $\sim$ 5 hours in the winter and $\sim$ 7.5 hours in the Summer, as those are the limits imposed by \poet's dome, which is not raised above Paranal's platform. We refer to \citet{santosPoETParanalSolar2025} for information on the dome location within the platform. It is also important to note that this limitation  on the zenith angle implies that, with the current setup, it is not possible to have a 24-hour coverage of the seeing in the Paranal observatory, as there is no overlap between the daytime measurements of the \shabar{} and the  nighttime seeing measurements from a DIMM \citep{1990A&A...227..294S} system from the European Southern Observatory (ESO).

\section{Data analysis and flow} \label{sect:shabar_data_analysis}

\begin{figure}[h]
      \centering
      \resizebox{\hsize}{!}{\includegraphics{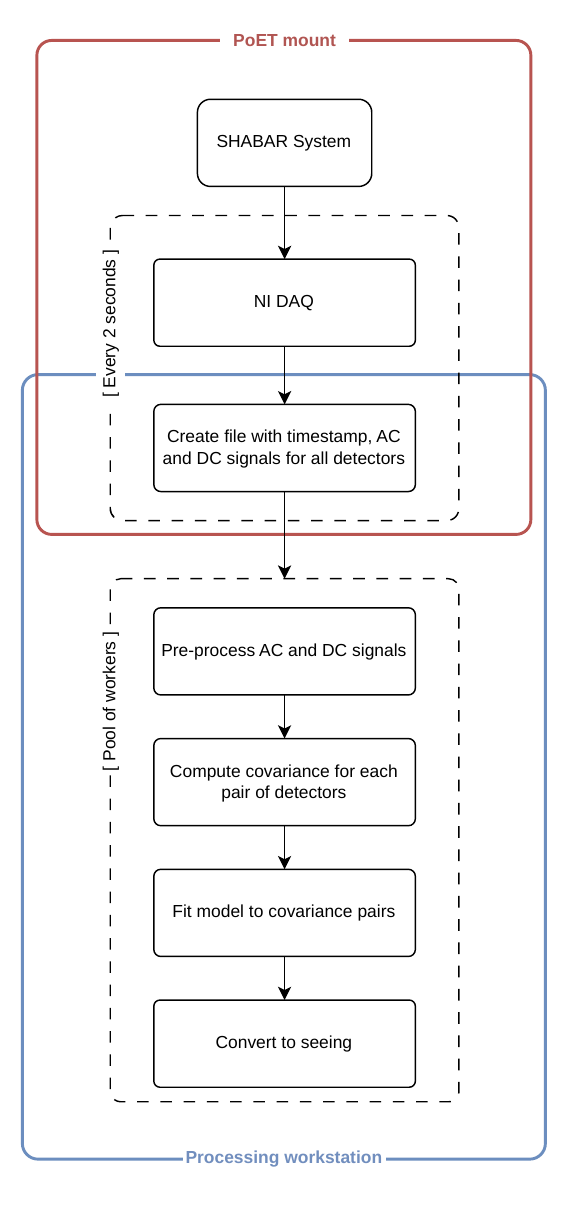}}
      \caption{Diagram of the data collection and reduction for the \shabar{} system installed in the \poet{} solar telescope. The data is collected through an industrial mini computer, attached to the telescope's mount, and then it is sent to a processing workstation through a network folder.}
      \label{Fig:poet_workflow}
   \end{figure}

   The \shabar{} system measures the scintillation of the incoming light, which is caused by the turbulence in the atmosphere. The scintillation is measured in the individual detectors, seen in Figure \ref{Fig:poet_picture}, which are then used to characterize the atmosphere at a given range of heights.  In this Section we shall describe the different stages of the data reduction process, with the full workflow schematized in Figure \ref{Fig:poet_workflow}.

   The \shabar{} processing workflow is divided between two different computers, as one is focused in data collection whilst the other in the data processing. In the \poet{} mount we have a mini-computer that is connected to a NI Data Acquisition (DAQ) system, collecting the AC and DC signals and storing them to disk. The collected data is then sent to a processing workstation, where the data is pre-processed and the correlations between detectors are computed. The correlations are then used to estimate the seeing, which is then made available to the observers through the dedicated interfaces.

   \subsection{Pre-processing of the AC and DC measurements} \label{Sec:shabar_preprocess}

      The AC and DC signals are collected over periods of 2 seconds at a frequency of 1000 Hz, for each detector in the system. Before those values are used to compute the correlation between detectors, we apply the following pre-processing steps:

      \begin{enumerate}
         \item Removal of samples near the dynamic range limits of the system ($\pm$ 9 V): If \poet{} is not pointing at the Sun, the recorded DC values will be close to the lower limit  of the ADC input range, and will not lead to meaningful values. As such, we reject samples where the average DC value for any of the detectors is smaller than -8.5 V.
         \item Removal of outliers: We apply a sigma clipping to the AC and DC values of each detector, removing samples that are more than 5 standard deviations away from the mean. This is done to remove spurious measurements that can be caused by transient events, such as cosmic rays or electronic noise. Furthermore, we also ensure that if the \textit{i}-th sample is rejected from a given detector, it will be rejected from all others, to ensure that the correlation computation is not affected by missing samples in one detector and not in the others. Lastly, if more than 75\% of the samples are rejected for the time series collected by any given detector, we discard the entire 2-second interval as it is likely that the measurements were contaminated by some unknown event.
         \item Conversion from voltage to Analogue to Digital Units (ADU): The signal amplification circuitry was designed such that the DC voltages will fall within the interval [-9, 9] V to match the ADC input dynamic range. When the Sun is close to the horizon the measured DC value is close to the lower limit of the ADC input range. However, as the Sun  rises, the DC value will increase, eventually crossing zero, reaching positive values. As the calculation of the correlation between detectors (see Sect. \ref{Eq:correlation_computation}) normalizes the signals by their respective mean value, DC values in the interval $]-1, 1[$ will lead to jumps in the correlation. To remove this issue, DC values are converted to ADU through Eq. \ref{Eq:voltage_to_counts} and only then the normalization step is executed. To ensure consistency in the analysis, the AC signal is also converted to ADUs, as per Eq. \ref{Eq:voltage_to_counts}:
      \end{enumerate}

      \begin{equation} \label{Eq:voltage_to_counts}
         C\ =\ (2^{n} - 1)\ *\ \frac{(S - V_{neg})}{\Delta\ V}
      \end{equation}
      
      \noindent where C represents the digital counts, S the measured signal, $n$ the number of bits in the ADC, $V_{neg}$ the maximum negative voltage of the system, and $\Delta\ V$ the dynamic range of the system.

   \subsection{Computation of the correlation between detectors} \label{Eq:correlation_computation}
      
      For each 2-second sample from the \shabar{} system we compute the correlation between every possible pair of detectors (we have 6 in total, resulting in 15 pairs). Before we compute the correlation, the AC and DC signals are normalized by their respective gains in each detector, as measured in the lab. The correlation between the \textit{i}-th and \textit{j}-th detectors, $\mathcal{C}_{i,j}$, is then given by:

      \begin{equation}
         \mathcal{C}_{i,j} = \frac{1}{N}\sum_{k=0}^{k=N}\frac{(AC_{i}^{k} - \overline{AC_{i}})\ *\ (AC_{j}^{k} - \overline{AC_{j}})}{\overline{DC_{i}}\ *\ \overline{DC_{j}}}  
      \end{equation}

      \noindent where $AC_{i}^{k}$ is the AC signal of the \textit{i}-th detector at the \textit{k}-th sample, $\overline{AC_{i}}$ is the mean AC signal of the \textit{i}-th detector, $\overline{DC_{i}}$ is the mean DC signal of the \textit{i}-th detector, and $N$ is the total number of samples in the 2-second interval (after pre-processing).

   \subsection{From correlation to seeing measurement}

      The conversion from inter-detector correlations to seeing measurements is based on Kolmogorov statistics, as prescribed in \citep{hardyAdaptiveOpticsAstronomical1998,hillDerivingCn2hScintillometera,sliepenSeeingMeasurementsAutonomous2010}. Below we present the main equations to model the correlation profile and estimate the seeing, referring the reader to the aforementioned works for a complete description and derivation of the formalism. Under this framework, the correlation as a function of the inter-detector separation is given by:

      \begin{equation} \label{Eq:correlation_model}
         \mathcal{C}(r)
         = 0.38\ *\
         \int_0^\infty
         C_n^2(h)
         K(h,r)\,dh
      \end{equation}

      \noindent where $r$ is the distance between each pair of scintillometers,  $C_{n}^{2}$ is the index of refraction structure parameter as a function of height, $h$ is the height at which we are evaluating our model, and $K(h,r)$ is a kernel given by:

      \begin{align} \label{eq:kernel}
         K(h,r)
         =
         \frac{32\pi h^2 \sec^3 \zeta}
         {\left(a+\alpha h\sec\zeta\right)^{7/3}}
         \,
         Q\!\left(
         \textit{s}
         \right)  \\
         \textit{s} = \frac{r}{a+\alpha h\sec\zeta}
      \end{align}
      
         \noindent where h is the height at which we are evaluating its value, \textit{r} is the distance between the scintillometers, \textit{a} is the lens's diameter of each photodiode, $\alpha$ is the angular diameter of the Sun at the time of the observation, $\zeta$ is the zenithal angle of the Sun, and $Q(s)$ is given by: 

      \begin{equation}     \label{eq:kernel_integral}
         Q(s)
         =
         \int_0^\infty
         \left[
         J_1(\pi f)
         \right]^2
         J_0(2\pi f s)\,
         f^{-2/3}\,df
      \end{equation}
      
      $J_0$ and $J_1$ are the zeroth- and first-order Bessel functions of the first kind. The computation of $\alpha$ and $\zeta$ is done through the \texttt{sunpy} \citep{communitySunPyProjectInteroperable2023} package, configured for the location of the PoET solar telescope. The integral of equation \ref{eq:kernel_integral} is pre-computed in a grid of values for \textit{s}, and then interpolated to the required value when computing the kernel. 

      \begin{figure}[h]
         \centering
         \resizebox{\hsize}{!}{\includegraphics{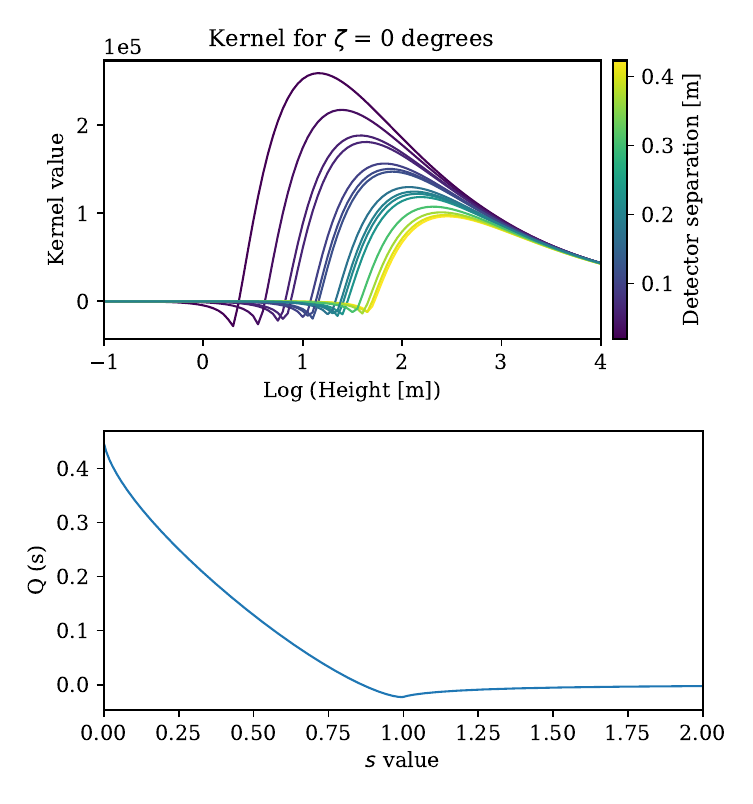}}
         \caption{\textbf{Top:} Height evolution of the kernel (Eq. \ref{eq:kernel}) for the separations of the detectors of the \shabar{} system, assuming a zenithal angle of 0 degrees. \textbf{Bottom:} Integral from Eq. \ref{eq:kernel_integral} evaluated at the region where the largest variation is present.}
         \label{Fig:kernel_and_integral}
      \end{figure}

      The top panel of Fig. \ref{Fig:kernel_and_integral} presents an evaluation of the kernel (Fig. \ref{eq:kernel}) for the different separations of the \shabar{} system, assuming a zenithal angle of 0 degrees.
      
      We can now approximate our $C_n^2$ profile as a combination of \textit{m} basis with a given amplitude, $a_i$, at fixed heights, $h_i$, such that:

      \begin{equation}
         C_n^2(h) = \sum_{i=0}^{m} a_i\delta(h-h_i)
      \end{equation}

      \noindent we can re-write the correlation model of Eq. \ref{Eq:correlation_model} as:

      \begin{equation}
         \mathcal{C}(r)
         = 0.38\ *\ \int_0^\infty \left( \sum_{i=0}^{m} a_i\delta(h-h_i) \right)\ K(h,r)\,dh
      \end{equation}

      \noindent which simplifies to:

      \begin{equation} \label{Eq:final_correlation_model}
            \mathcal{C}(r)
         = 0.38\ *\ \sum_{i=0}^{m} a_i K(h_i,r)
      \end{equation}

      \noindent We can then fit the model of Eq. \ref{Eq:final_correlation_model} to the measured correlation values, using a least-squares minimization, to estimate the amplitudes of the $C_n^2$ profile at the fixed heights. We accomplish this through a Nelder-Mead optimization algorithm, implemented in the \texttt{scipy} \citep{virtanenSciPyFundamentalAlgorithms2020} package, applied to the following metric:

      \begin{align}
         E\ = E_m + \Lambda\ *\ E_r \\
         E_m\ =\ \frac{1}{N}\sum_{i=0}^{N} \left( \mathcal{C}_{meas}(r_i) - \mathcal{C}_{model}(r_i) \right)^2 \\
         E_r = \frac{1}{2(N_h-1)} \sum_{i=1}^{N_h} \left( \log C_n^2(h_i) - \log C_n^2(h_{i-1})\right)^2
      \end{align}
      \noindent with $E_r$ introducing a regularization term, ensuring a smooth slope of the $C_n^2$ in a logarithmic parameter space. In this formalism, the $\Lambda$ scaling parameter ensures that the regularization term is not dominating the optimization, but it is still large enough to ensure a smooth solution. We set $\Lambda$ to $10^{-7}$, which was found to be a good compromise between the two terms, ensuring that the optimization converges to a stable solution without over-smoothing the height profile. Lastly, the optimization is performed in logarithmic space, ensuring that the $C_n^2$ values are always positive. In total, we use 20 steps that are uniformly spaced in log height, with the highest one being set at 10 km. The initial conditions for the optimization are set to a constant value of $C_n^2\ =\ 10^{-14}$ across the full height interval. Furthermore, after the optimization is complete we start it again, using the previous results as the initial conditions, ensuring that the optimization converges to a stable solution.

   \subsection{Conversion to seeing measurements} \label{Sect:seeing_computation_from_cn}
      
      For each individual sample we apply the recipe of Sect. \ref{sect:shabar_data_analysis}, computing a high-cadence time series of the $C_n^2$ profile, which then needs to be transformed into a seeing measurement. We start by computing the Fried parameter, $r_0$, through Eq. 3.51 of \citet{hardyAdaptiveOpticsAstronomical1998}:

      \begin{equation}
         r_0 = \left(  0.423\ k^2\ sec(\zeta) \int\ C_n^2(h)\ dh  \right) ^ {-3/5}
      \end{equation}

      \noindent, which is then transformed into seeing ($\epsilon_0$) measurement:

      \begin{equation}
         \epsilon_0\ \approx\ \frac{\lambda}{r_0}
      \end{equation}

      \noindent where $\lambda$ is the operating wavelength of the \shabar{} system (565 nm for our instrument), defined by a narrow bandpass filter embedded on the photodiode. Lastly, we apply a moving average with a window of 60 seconds to the resulting seeing time series, removing high-frequency variations in the measurements and leading to more stable results. In appendix \ref{App:injection_recovery} we present a series of injection-recovery tests, where we inject a known $C_n^2$ profile into our pipeline, add random noise, and then recover it through the formalism described in this Section. We find that the recovered seeing measurements are in good agreement with the injected ones, validating our data analysis and processing workflow.

\section{Daytime seeing variations in Paranal} \label{Sec:paranal_daytime_seeing}

   In this Section we present the daytime seeing measurements collected during the first weeks of operations of \poet. The data shown in this Section was collected between April 27th and July 11th, 2026, with a total of 37 days of observations. Due to multiple technical interventions, downtime of \espresso{}, and downtime of the \poet{} telescope, the data is not continuous, with some days only having a few hours of observations.
   
   \subsection{Intra-day variation}
\begin{figure}[h]
         \centering
         \resizebox{\hsize}{!}{\includegraphics{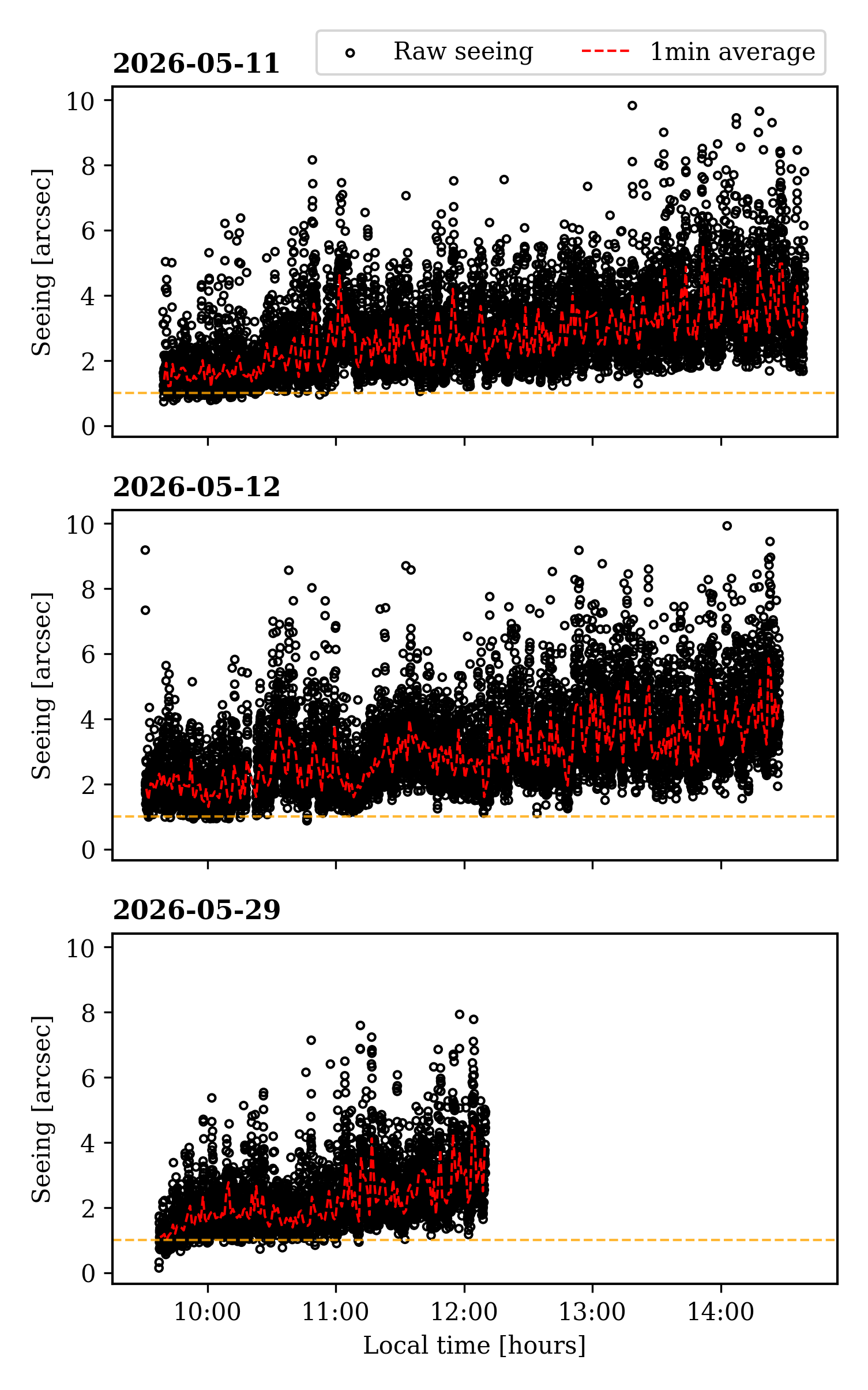}}
         \caption{Intra-day seeing measurements, as measured through the \shabar{} instrument with a cadence of 2 seconds. In red, we overplot a 1-minute average of the measurements, to remove the short-term variability in the data, whilst the orange line represents a 1 arcsecond seeing. Each row presents one day worth of observations, as indicated in the upper left corner of each panel.}
         \label{Fig:intra_day_seeing}
      \end{figure}

      Our first analysis of daytime seeing through our \shabar{} instrument is done through intra-day variations, focusing on the seeing variability over a few hours of observations, both in the raw  measurements (2 second cadence) from the detector and through a 1 min-average of those values. In Figure \ref{Fig:intra_day_seeing} we show the results for three days of \poet{} operations, with one of them limited in time due to operational constraints.

      Through this preliminary analysis we find that the daytime seeing in the Paranal observatory can reach  the 1 arcsecond level for small periods of time during the morning. The raw measurements that are shown in this Figure reveal some seeing variability in short time-scales, similarly to what was previously observed in our \shabar{} comissioning run \citep{wehbeSeeingMeasurementDevice2026}. If we bin the seeing measurements into 1 minute interval we get a more stable measurement, overcoming very short term variability, as shown in the red dashed line of Figure \ref{Fig:intra_day_seeing}, and matching the typical 1 minute cadence of  \espresso{} observations from \poet{} (accounting for exposure and CCD readout time).

   \subsection{Inter-day variation over the first month of observations}

   \begin{figure*}[h!]
      \sidecaption
      \includegraphics[width=12cm]{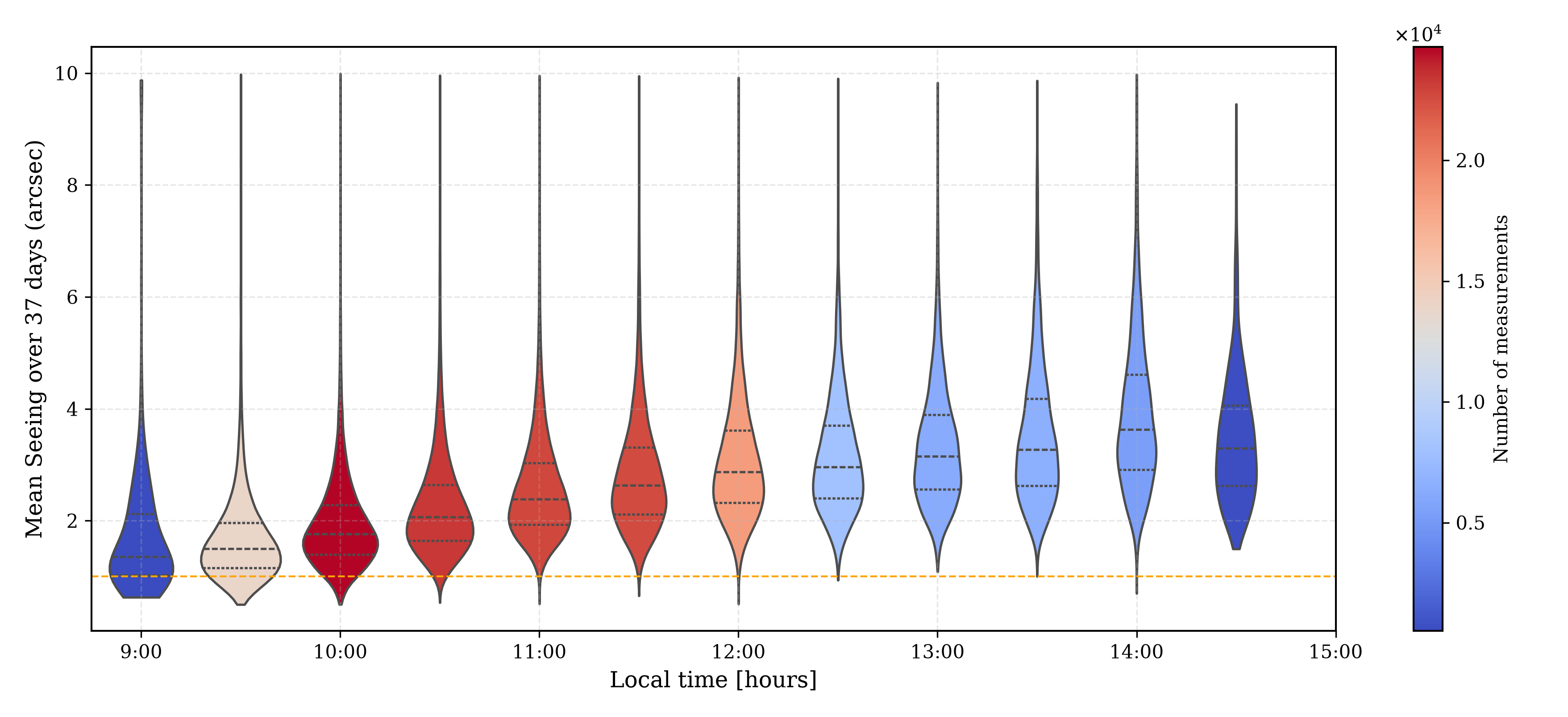}
      \caption{Mean seeing over 30 minute bins, as measured over 37 days of \poet{} operations. The violin plots show the distribution of the seeing measurements, with the dashed lines showing the 25th and 75th percentiles of the distribution, and the dashed horizontal line representing a seeing of 1 arcsecond. The color of each bin represents the number of measurements that fall inside it, as specified from the colorbar.}
      \label{Fig:inter_day_variation}
   \end{figure*}
   
      If we now focus on a larger temporal window, we can evaluate how the seeing evolves over multiple days. Using the first weeks of measurements from our \shabar{} system we recover a median seeing of 2.37 arcseconds over all days in our sample, well in agreement with the reported 2.65 arcseconds reported by \citet{griffithsComparisonNextgenerationTurbulence2024}. A direct comparison between the two values is not possible, as the measurements of \citet{griffithsComparisonNextgenerationTurbulence2024} were collected in a different period of the year, with a different instrument, and in a different location in Paranal's platform. However, the agreement between the two campaigns reassures us that our measurements are in line with the expected values for the daytime seeing in Paranal.
      
      To assess the daily evolution and stability of the seeing conditions we bin the seeing values in 30 minute intervals, in local time, and compare their distributions over the day. The results are shown in Fig. \ref{Fig:inter_day_variation}, where we can see that the median seeing increases over the day, likely reflecting the ground temperature increase, and thus the increase of ground layer turbulence.

      Particularly at the start of the day we find that the seeing often reaches values smaller than 3 arcseconds, with some days reaching values smaller than 1 arcsecond before 10:00 AM, local Chilean time, allowing solar observations with \poet's smallest on-sky aperture of 1 arcsecond. We also find that there are large outliers in the selected intervals, which can be caused by transient events that are caught due to the high-sampling of the instrument, leading to the high-seeing excursions that can be seen across the day.

\section{Conclusions and discussion} \label{Sec:conclusions}

   In this manuscript we have presented the software architecture and the first daytime seeing measurements that were collected from a \shabar{} system that is coupled with the recently comissioned \poet{} solar telescope, installed at ESO's VLT platform in Atacama. The seeing measurements that are collected by this system play a pivotal role in the observational strategy of this solar telescope, as it employs fibers ranging between 1 and 55 arcseconds to observe resolved solar regions. As such, the usage of the smaller apertures hinges upon good observational conditions, which are assessed through the real-time measurements of the piggyback \shabar{} system.

   We present results from the first weeks of observations after \poet's comissioning, focusing on intra- and inter-day seeing variations. At the level of a single day we often find that seeing in the early morning (before $\sim$ 10:00, local winter time) can reach the 1-arcsecond level. Over the period of a full day of observations we find, as expected, an increase of the seeing. This is likely a consequence of the increase of the ground temperature over the day, leading to stronger turbulence in the lower layers of the atmosphere, which are the ones that most affect our telescope, as the dome is at the ground level. Through 37 days of observations we also find that the same tendency is observed across the day, with the median seeing increasing over the day, and reaching values larger than 3 arcseconds in the afternoon. As such, and taking this small temporal window of observations, the Paranal observatory boasts the sufficient atmospheric conditions for operations of daytime telescopes, especially in cooler months. 
   
   Lastly, it is however important to note that our measurements were collected in late Autumn and early winter in Chile, leading to lower temperatures and, consequently, a smaller increase of temperature over the day. As such, it is fully expected that seasonal changes will greatly impact the seeing conditions, possibly leading to larger seeing values than those reported here. We leave for future work a seasonal analysis of the daytime seeing in the VLT platform, coupled with a cross-match between the daytime measurements collected by our \shabar{} system and the nighttime ones collected through the DIMM system, as provided by ESO.

\begin{acknowledgements}
   This work was funded by the European Union (ERC, FIERCE, 101052347). Views and opinions expressed are however those of the author(s) only and do not necessarily reflect those of the European Union or the European Research Council. Neither the European Union nor the granting authority can be held responsible for them. This work was also supported by Fundação para a Ciência e a Tecnologia (FCT) through national funds under the research grant UID/04434/2025 (DOI 10.54499/UID/04434/2025). J. H. C. M. acknowledges further support from the Portuguese Space Agency through projects e-CHEOPS (PEA: 4000142255) and PLATO (PEA: 4000133026, 4000140773, and 4000124770), both funded by ESA/PRODEX
\end{acknowledgements}

%

\bibliographystyle{aa}
\bibliography{paranal_seeing_bib}

\begin{appendix}

\section{Validation through injection-recovery tests} \label{App:injection_recovery}

In order to test our model ability to accurately  retrieve atmospheric seeing we carried out injection-recovery tests. This was done through the following steps:

\begin{itemize}
   \item We generate a synthetic $Cn^2(h)$ profile with fixed coefficients, based on a standard Huffnagel 5-7 model \citep[e.g., see Table 3.1 of][for a description of the model]{hardyAdaptiveOpticsAstronomical1998}.
   \item Through Eq. \ref{Eq:correlation_model} the synthetic profile is used to construct a set of inter-detector correlation profiles, assuming the detector separation of our system. 
   \item Random noise is added by drawing realizations of a normal distribution centered on the synthetic correlation values, with a standard deviation equal to 10 \% of their value. 
   \item The formalism of Sect. \ref{Sect:seeing_computation_from_cn} is used to compute the expected seeing value from the injected $Cn^2(h)$ profile.
\end{itemize}

\begin{figure}[h]
   \centering
   \resizebox{\hsize}{!}{\includegraphics{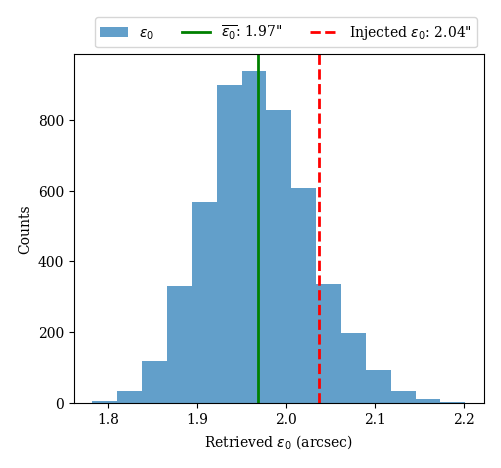}}
   \caption{Histogram of the retrieved seeing values, in the presence of a 10 \% noise level on the correlation measurements. The injected seeing ($\epsilon_0\ =$ 2.04") is displayed with a dashed red line, and the mean retrieved seeing ($\overline{\epsilon_0}\ =$ 1.97") is shown in a green line. }
   \label{Fig:seeing_inj_recov}
\end{figure}

This procedure is repeated 5000 times, generating independent realizations of the synthetic correlation profiles. The \shabar{} data analysis pipeline is then used to estimate the atmospheric seeing, with Fig. \ref{Fig:seeing_inj_recov} showing the results.

Trough this comparison we find that our seeing computation is robust, providing a mean seeing value that is consistent with the injected one, with differences at the level of the 0.1 arcseconds, 1 order of magnitude smaller than \poet{}'s smallest fiber. 

\end{appendix}

\end{document}